\documentclass[twocolumn,showpacs,pra]{revtex4}

\usepackage{graphicx}

\begin{document}
\title{Systematic study of the decay rates of antiprotonic helium states} 

\author{H. Yamaguchi,  R. S. Hayano, T. Ishikawa, J. Sakaguchi, and E. Widmann}
\affiliation{Department of Physics, University of Tokyo, 7-3-1 Hongo, Bunkyo-ku, Tokyo 113-0033, Japan}

\author{J. Eades and M. Hori}
\affiliation{CERN, CH-1211 Geneva 23, Switzerland}

\author{H. A. Torii }
\affiliation{Institute of Physics, University of Tokyo, Komaba, Meguro-ku,
Tokyo 153-8902, Japan }

\author{B. Juh\'{a}sz}
\affiliation{Institute of Nuclear Research of the Hungarian Academy of Sciences, H-4001 Debrecen, Hungary}

\author{D. Horv\'{a}th}
\affiliation{KFKI Research Institute For Particle and Nuclear Physics, H-1525 Budapest, Hungary}

\author{T. Yamazaki}
\affiliation{
RI Beam Science Laboratory, RIKEN, Wako, Saitama 351-0198, Japan
}
\date{\today}

\newcommand{\pbar}{\ensuremath{\overline{p}}}
\newcommand{\pbhe}{\pbar He$^+$}
\newcommand{\pbhef}{\mbox{\pbar$^4$He$^+$}}
\newcommand{\pbhet}{\mbox{\pbar$^3$He$^+$}}

\begin{abstract}

A systematic study 
of the decay rates of antiprotonic helium 
(\pbhef\ and \pbhet) at CERN AD (Antiproton Decelerator)
has been made by a laser spectroscopic method.
The decay rates of some of its short-lived states, 
namely those for which the Auger rates $\gamma_{\mathrm{A}}$ are much larger 
than their radiative decay rates ($\gamma_{\mathrm{rad}} \sim 1$ $\mu$s$^{-1}$),
 were determined from 
the time distributions of the antiproton annihilation signals 
induced by laser beams,
and the widths of the atomic resonance lines.  
The magnitude of the decay rates, especially their relation 
with the transition multipolarity, is discussed and compared
with theoretical calculations.
\end{abstract}


\maketitle

\section{Introduction}
\label{sec:intro}
The antiprotonic helium atom (\pbhe) is an exotic three-body system 
consisting of an antiproton, an electron, and a helium nucleus 
\cite{Iwasaki:91,Yamazaki:93,Yamazaki:02}.
Detailed studies have been done for 
its metastable-state region, 
where the principal quantum number $n$ and angular momentum $l$ 
are both around 38, and where lifetimes against 
annihilation can be as long as several microseconds.
Recently new CPT-violation limits 
on the antiproton mass and charge were determined
\cite{Hori:03} at CERN AD (Antiproton Decelerator)
using  a laser spectroscopic method
to measure some transition energies of the \pbhe\ atom to a precision of 
50--200 ppb. 
The measured transition energies were 
compared with two independent variational 
three-body calculations with QED corrections 
by Korobov \cite{Korobov:96,Korobov:97a,Korobov:03} and 
by Kino \cite{Kino:01,Kino:03}. 

In addition to their level energies $E_r$,
an important property of the states of this exotic atom is 
their decay rate $\gamma$,
which forms the imaginary part of the complex energy,
\begin{equation}
E = E_r -i\frac{\gamma/2\pi}{2}.   
\label{complex_energy} 
\end{equation}
In particular, the Auger rates $\gamma_A$, which are dominant in the total 
decay rate for some \pbhe\ states,
have been calculated by many theorists 
\cite{Russell:70c,Yamazaki:92,Korobov:97b,Kartavtsev:00}
over a period of more than 30 years.
The latest calculations 
with CCR (Complex Coordinate Rotation) method  by 
Korobov \cite{Korobov:03} and Kino \cite{Kino:03}
yield  both real and imaginary parts of the complex energy
at the same time, and the both parts are expected to have
absolute precisions of the same order.
An independent check on the validity of these three-body calculations 
can therefore be obtained by measuring the decay rates of 
Auger-dominant states.

Although the CPT-violation limits \cite{Hori:03} were obtained by 
the level-energy measurements with 
high relative precisions of 50--200 ppb,
discussions of the decay rates with 
such high accuracy are 
difficult both experimentally and theoretically.
The experimental difficulty originates in
the finite response time in the detection of the antiproton annihilations,
and in the broadening of the atomic spectral linewidth 
from various sources (laser linewidth, Doppler width and so on).
Theoretically, the  difficulty lies in
representing continuum-coupled discrete states 
with a limited set of wave functions,
and in the small magnitude of the Auger widths (order of MHz--GHz)
compared to the transition energies (0.4--1 PHz, from visible light to UV).
However, due to the smallness of the Auger widths,
the absolute precisions of  
both  experimental and theoretical decay rates
can be as good as those of the level energies 
for some states,
and their measurements 
would be meaningful
for the verification of the CCR calculations.

A prominent and important  feature of the theoretically calculated Auger rates
is their drastic dependence on the
lowest possible transition multipolarity $L$, which is equal to the 
minimum angular momentum carried away by the Auger electron
\cite{Russell:70c,Yamazaki:92}.
The theoretical rates are approximated by a rough estimation, 

\begin{equation}
\gamma_{A} \sim  10^{17-3L}~{\rm s}^{-1}~ {\rm for}~L=2,3,4. \label{auger_systematics}
\end{equation}
States with $L \le 3$ have radiative rates 
of  $10^5$--$10^6$ s$^{-1}$. They are consequently Auger-dominated,  
and we refer to them as ``short-lived'' states.
For states $L > 3$, fast Auger decays are suppressed, and
the atom de-excites only radiatively. 
These are the states which we call ``metastable''.

By 2001, we had studied the decay rates of eight \pbhef\ states \cite{Hori:98,Yamaguchi:02}.
Although most of them have decay rates in accordance with the 
above approximate Eq.~(\ref{auger_systematics}), 
we found two exceptions.
One was the ($n$, $l$)=(37, 33) state, which is possibly affected by  
an electron-excited configuration $(32, 31)_{\bar{p}} \otimes (3d)_e$ 
having a nearby level energy \cite{Kartavtsev:00}.
Another was the (32, 31) state,
the decay rate of which
our low-density measurement with the RFQ decelerator 
\cite{Hori:03,Lombardi:01} revealed to be 
greatly enhanced by collisions.
We call these decay-rate discrepancies ``anomalies''
\cite{Yamaguchi:02}. 

\section{Measurement of decay rates}
\label{sec:analysis}

In 2001--2003, we studied decay rates of ten 
states of \pbhe, including seven states of \pbhet.
A detailed description of the experimental procedure using laser spectroscopy 
can be found in Refs. \cite{Morita:93,Morita:94,Hori:01}.

Antiproton pulses of duration $\sim$ 300 ns, each of which 
contained some $10^7$ particles were brought to rest in a helium gas target
at  a temperature of 5.5 K and pressures between 20 and 200 kPa. 
Charged pions produced  by their subsequent annihilation were detected
by \v{C}erenkov counters, time spectra of these pions 
being recorded as analog pulses (``ADATS'', for  Analog Delayed 
Annihilation Time Spectrum), by a digital oscilloscope \cite{Niestroj:96}.

When we applied  a laser pulse tuned to a wavelength of 
a metastable to short-lived transition, a sharp peak appeared in 
the ADATS, the intensity (area) of which was 
 proportional to the number of antiproton annihilations induced 
by the laser pulse. 
Spectral line profiles could then be obtained by plotting the peak 
intensity against the laser wavelength as this was scanned through 
the central transition wavelength. 

Three methods were used to deduce decay rates
according to their magnitude.

\subsection{Decay rates faster than 6 ns$^{-1}$ (wide resonances)}
When the decay rate is fast enough, a transition 
to the state is observed as 
a wide resonance having a natural width larger 
than the laser linewidth 
($\sim$ 1 GHz at ``high resolution'' mode, see below).
Fig.~\ref{fig:wide} shows four $(n, l) \to (n+1, l-1)$ type 
resonances newly measured in this work.
\begin{figure}[!htb]
\centerline{\includegraphics[width=.9\linewidth,height=.6\linewidth]{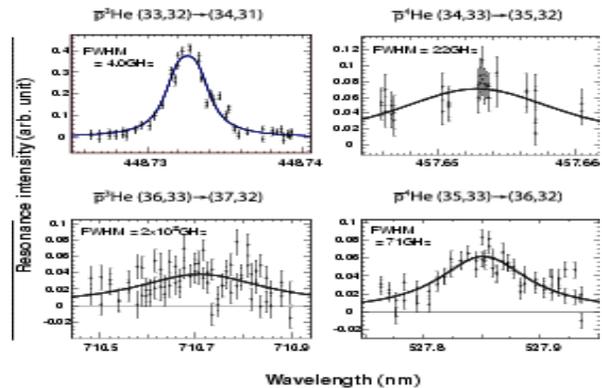}}
\caption{\label{fig:wide} Wide resonances scanned for the Auger rate deduction.
Their FWHM (full width at half maximum) were obtained by 
fitting with Lorentz functions.
Statistical errors are shown.} 
\end{figure}
Such transitions are called ``unfavored'',
since they have typically ten times smaller transition dipole moments
than $(n, l) \to (n-1, l-1)$ (or  ``favored'') transitions.
The measurements were performed at a specific target atomic density of 
$3\times 10^{20}$ cm$^{-3}$ (6 K, 20--30 kPa),
since for most states (\cite{Hori:98,Yamaguchi:02}) 
no systematic density variation of the decay rates
larger than the experimental error was observed.

The dye laser  used in our laser system (Lambda Physik Inc. Scanmate 2E) 
could be operated in ``broadband'' or ``high resolution'' modes.
In the ``broadband'' mode, the power was high 
(a few 10 mJ/shot) but the 
linewidth was several GHz, 
while in the ``high resolution'' mode, 
the laser frequency linewidth was severely 
constrained by an intracavity etalon
to about 1 GHz.
For some of the unfavored transitions shown in Fig.~\ref{fig:wide}, 
the laser was operated in the broadband mode to fulfill the requirement of 
high power for the unfavored resonances.
It is difficult to know the exact linewidth during the resonance scans.
By investigating the spectra broadening measured by Fizeau interferometers,
the linewidth could be determined as  4--10 GHz, except for the
448-nm case for which the laser was operated in the high resolution mode
with a linewidth of 0.7--1.3 GHz.

The analysis procedure for each of these states was the following.
The resonance spectrum was fitted with 
two identical Voigt functions (convolutions of a Gaussian and a Lorentzian), 
separated by a fixed frequency splitting.
This splitting was introduced to represent 
the hyperfine splitting \cite{Widmann:97,Widmann:02},
caused by the coupling of the electron spin and
the \pbar~orbital magnetic moment.
For the hyperfine splitting, theoretical values by Bakalov and Korobov 
\cite{Bakalov:98} 
were used, or  (2 $\pm$ 0.5) GHz assumed for uncalculated states.
The width of the Gaussian component was fixed at the laser linewidth plus
Doppler broadening,
and the Lorentzian width was obtained from the fitting procedure. 
The Lorentzian width was considered to be
approximately equal to the natural width $\Delta \nu_{\text{nat}}$
because the other factors such as 
the collisional and power broadening were 
estimated to be much smaller than 1 GHz.
Finally, 
$\Delta \nu_{\text{nat}}$ was converted to the decay rates $\gamma$ by 
\begin{equation}
\gamma = 2 \pi \Delta \nu_{\text{nat}}.
\label{natural_width} 
\end{equation}
The errors were estimated mainly from
the statistical fluctuations, uncertainty of the laser linewidths and 
hyperfine splittings.

\subsection{Decay rates of about 0.2 ns$^{-1}$ or slower (long tail peaks)}
If the decay rates are slow enough,  
antiprotons which annihilate from the Auger-dominated state
do so after a measurable additional delay. 
We can then obtain the decay rates directly by fitting
the ``tails'' of the laser resonance peaks with exponential 
functions. 
The effect of the fall time of the Hamamatsu Photonics 
R5505GX-ASSYII photomultiplier tubes (about 4 ns)
in prolonging these tails was corrected for in the 
fitting procedure. 
This method is applicable for decay
rates up to about 0.1--0.2 ns$^{-1}$,  depending on 
the performance of the photomultiplier tubes.
\begin{figure}[!htb]
\centerline{\includegraphics[width=.9\linewidth]{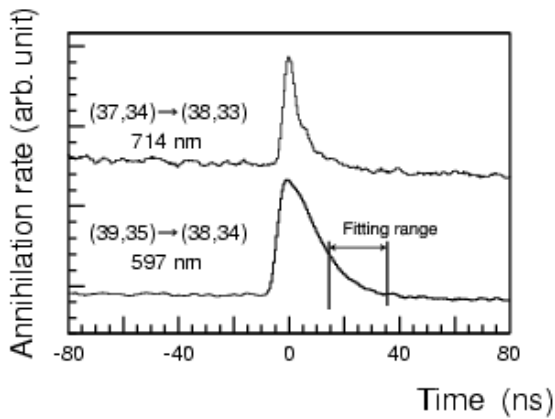}}
\caption{\label{fig:tail} Prolonged laser resonance peak (lower) 
and unprolonged one (upper). 
} 
\end{figure}

In this work, only the (34,32) state of \pbhef\ could 
be considered to fall in this category.
This state was measured also in the previous work 
\cite{Yamaguchi:02};
in the present experiments we obtained a more reliable result
by using a fast-response photomultiplier tube.

\subsection{Intermediate rates} 
\label{sec:case_inter} 
For the states that do not show either wide resonances 
 or prolongation of laser peaks,
we cannot measure their decay rates directly.
The daughter states of an unfavored 
[the $(35, 33) \to (36, 32)$ of \pbhet]
and five favored resonances
[to the (37, 33), (35, 32), (33, 31), and (31, 30) of \pbhet,
and the (39, 34) of \pbhef\ from each corresponding $(n+1, l+1)$ state] 
come under this category.
In this case, we concluded that their decay rates are between 
the two experimental limits.
The lower limit (0.1 ns$^{-1}$) is determined mainly 
by the instrumental prolongation
effect of the laser peak, as mentioned above.
The upper limit is determined by the laser linewidth.  
It is 6 ns$^{-1}$ (corresponding to 1 GHz 
width) for the high resolution scans,
and 20 ns$^{-1}$  for the broadband scan.
The latter value was obtained by assuming that the 
laser bandwidth in broadband mode is wider than 4 GHz.

\section{Results and discussion}
\label{sec:results}

\begin{figure*}[!htb]
\centerline{\includegraphics[width=.9\linewidth]{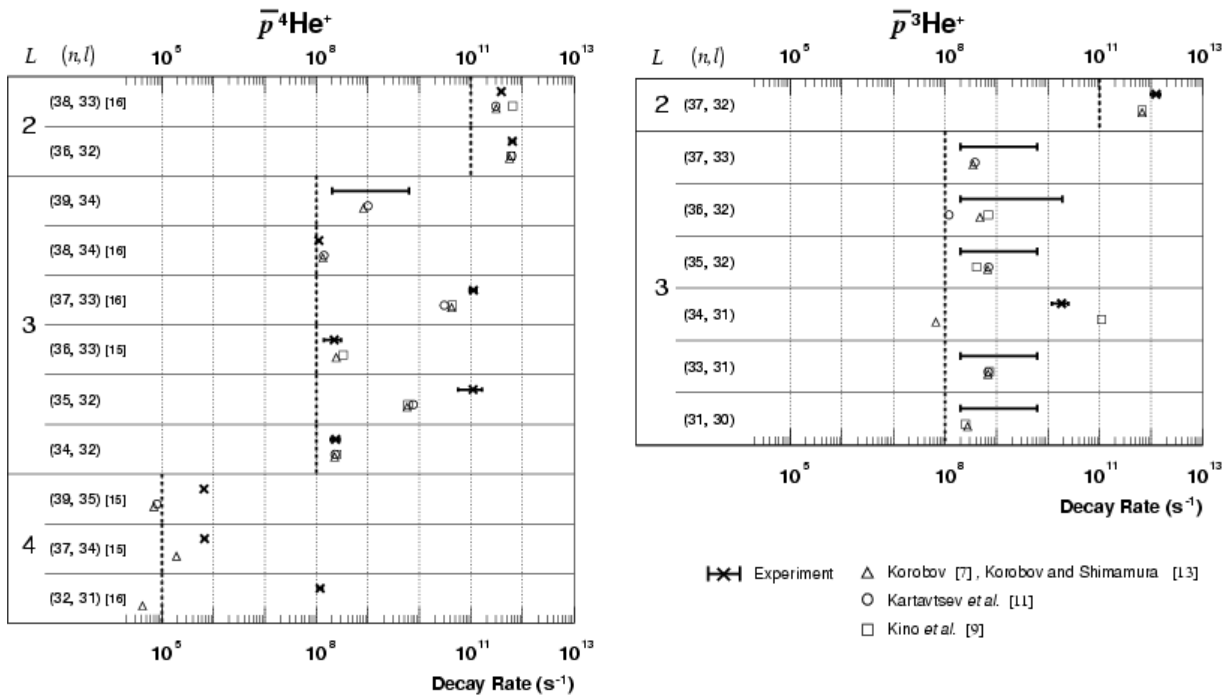}}
\caption{\label{fig:auger} Experiment-theory comparison of the decay rate in logarithmic scale.
All the data including past measurements \cite{Hori:98,Yamaguchi:02} are shown.
The bold vertical dotted lines show the typical Auger rates 
estimated by  Eq.~\ref{auger_systematics}.
The experimental points without crossed markers show
that the rates are between two experimental limits,
and not directly determined
(see the case of Sec.~\ref{sec:case_inter}).
Some of the experimental error bars are too small to see. 
} 
\end{figure*}

Fig. \ref{fig:auger} and Table \ref{tab:auger} show 
the results of our decay rate measurements
for all the eighteen (including  eight already published) states.
\begin{table*}
\caption{\label{tab:auger} Experimental decay rates and theoretically 
calculated Auger decay rates \cite{Korobov:97b,Kartavtsev:00,Korobov:03,Kino:03} 
in s$^{-1}$. $L$ is
the minimum transition multipolarity of Auger decay.
All the data including past measurements \cite{Hori:98,Yamaguchi:02} are shown.
For the  (39, 35) and (37, 34) states of \pbhef, the
radiative decay is dominant and the experimental decay rates
cannot directly be compared with theoretical Auger rates.
Some of the experimental values are not directly determined 
and only the possible ranges are shown
(see the case of Sec.~\ref{sec:case_inter}). 
}
\begin{ruledtabular}
\begin{tabular}{rcrcllp{2em}rclrclrclrcl}
\multicolumn{1}{c}{State} &$L$& \multicolumn{4}{c}{Experimental values} & 
& \multicolumn{3}{c}{KRB-SMM \cite{Korobov:97b}} & \multicolumn{3}{c}{KRT \cite{Kartavtsev:00}} & \multicolumn{3}{c}{KRB \cite{Korobov:03}} & \multicolumn{3}{c}{KNO \cite{Kino:03}}\\
\colrule
\pbhef
(38, 33)& 2& $(3.85 \pm 0.26)$&$\times$&$10^{11}$ &\cite{Yamaguchi:02} && 
$3.1$&$\times$&$10^{11}$ &$3$&$\times$&$10^{11}$ & 
$3.07$&$\times$&$ 10^{11}$ &$6.45$&$\times$&$10^{11}$  \\

(36 ,32)& 2& $(6.35 \pm 0.65)$&$\times$&$ 10^{11}$ &&& 
$5.8$&$\times$&$ 10^{11}$ &$6.1$&$\times$&$ 10^{11}$ &
$5.59$&$\times$&$ 10^{11}$ &$5.95$&$\times$&$ 10^{11}$ \\

(39 ,34)& 3& $(0.1$--$6)$&$\times$&$ 10^{9}$ &&& 
$7.7$&$\times$&$ 10^{8}$ &$1$&$\times$&$ 10^{9}$ & 
$8.21$&$\times$&$ 10^{8}$ \\

(38, 34)& 3& $(1.11 \pm 0.07)$&$\times$&$ 10^{8}$ &\cite{Yamaguchi:02} && 
$1.3$&$\times$&$ 10^{8}$ &$1.4$&$\times$&$ 10^{8}$ & 
$1.344$&$\times$&$ 10^{8}$\\

(37, 33)& 3& $(1.11 \pm 0.16)$&$\times$&$ 10^{11}$ &\cite{Yamaguchi:02} && 
$5.7$&$\times$&$ 10^{9}$  &$3$&$\times$&$ 10^{10}$ & 
$4.21$&$\times$&$ 10^{10}$&$4.38$&$\times$&$ 10^{10}$ \\

(36, 33)& 3& $(2.2 \pm 0.8)$&$\times$&$ 10^{8}$ &\cite{Hori:98}&& 
$2.4 $&$\times$&$ 10^{8}$ & &&& 
$2.42$&$\times$&$ 10^{8}$ & $3.31$&$\times$&$ 10^{8}$ \\

(35, 32)& 3& $(1.10 \pm 0.53)$&$\times$&$ 10^{11}$ &&& 
$3.7$&$\times$&$ 10^{9}$ &$7.5$&$\times$&$ 10^{9}$ & 
$5.76$&$\times$&$ 10^{9}$ &$6.04$&$\times$&$ 10^{9}$ \\

(34, 32)& 3& $(2.36 \pm 0.47)$&$\times$&$ 10^{8}$ & && 
$2.2$&$\times$&$ 10^{8}$ &$2.3$&$\times$&$ 10^{8}$ & 
$2.260$&$\times$&$ 10^{8}$ &$2.5$&$\times$&$ 10^{8}$\\

(39, 35)& 4 (rad.) & $(6.5 \pm 0.2)$&$\times$&$ 10^{5}$ &\cite{Hori:98} && 
$7.0 $&$\times$&$ 10^{4}$  & $8 $&$\times$&$ 10^4$\\

(37, 34)& 4 (rad.)& $(6.7 \pm 0.5)$&$\times$&$ 10^{5}$ &\cite{Hori:98} && 
$1.9 $&$\times$&$ 10^{5}$  & \\

(32, 31)& 4& $(1.17 \pm 0.12)$&$\times$&$ 10^{8}$ &\cite{Yamaguchi:02} && 
$6.1$&$\times$&$ 10^{5}$  &  &&& 
$4.2$&$\times$&$ 10^{4}$\\
\hline
\pbhet
(37, 32)& 2& $(1.24 \pm 0.26)$&$\times$&$ 10^{12}$ &&& 
$6.6$&$\times$&$ 10^{11}$  &  &&& 
$6.71$&$\times$&$ 10^{11}$ &$6.74$&$\times$&$ 10^{11}$\\

(37, 33)& 3& $(0.1$--$6)$&$\times$&$ 10^{9}$ &&& 
$3.3$&$\times$&$ 10^{8}$  & $3.8$&$\times$&$ 10^{8}$ & 
$3.52$&$\times$&$ 10^{8}$\\

(36, 32)& 3& $(0.1$--$20)$&$\times$&$ 10^{9}$ &&& 
$1.1$&$\times$&$ 10^{12}$  & $1.2$&$\times$&$ 10^{8}$ & 
$4.8$&$\times$&$ 10^{8}$ & $7.0$&$\times$&$ 10^{8}$\\

(35, 32)& 3& $(0.1$--$6)$&$\times$&$ 10^{9}$ &&& 
$6.8$&$\times$&$ 10^{8}$  & $7$&$\times$&$ 10^{8}$ & 
$6.76$&$\times$&$ 10^{8}$ & $4.1$&$\times$&$ 10^{8}$\\

(34, 31)& 3& $(1.84 \pm 0.65)$&$\times$&$ 10^{10}$ &&& 
$3.5$&$\times$&$ 10^{10}$  &  &&& 
$6.7$&$\times$&$ 10^{7}$ & $1.09$&$\times$&$ 10^{11}$\\

(33, 31)& 3& $(0.1$--$6)$&$\times$&$ 10^{9}$ &&& 
$6.9$&$\times$&$ 10^{8}$  & $6.9$&$\times$&$ 10^{8}$ & 
$6.85$&$\times$&$ 10^{8}$ & $7.4$&$\times$&$ 10^{8}$\\

(31, 30)& 3& $(0.1$--$6)$&$\times$&$ 10^{9}$ &&&
&&& &&& 
$2.76$&$\times$&$ 10^{8}$ & $2.5$&$\times$&$ 10^{8}$\\
\end{tabular}
\end{ruledtabular}
\end{table*}
Theoretical values by  Korobov and Shimamura \cite{Korobov:97b}, 
Kartavtsev with co-workers \cite{Kartavtsev:00}, 
and the latest calculations by Korobov \cite{Korobov:03} 
and Kino \cite{Kino:03} are also shown.

\begin{figure}[!ht]
\begin{center}
\includegraphics[width=.9\linewidth]{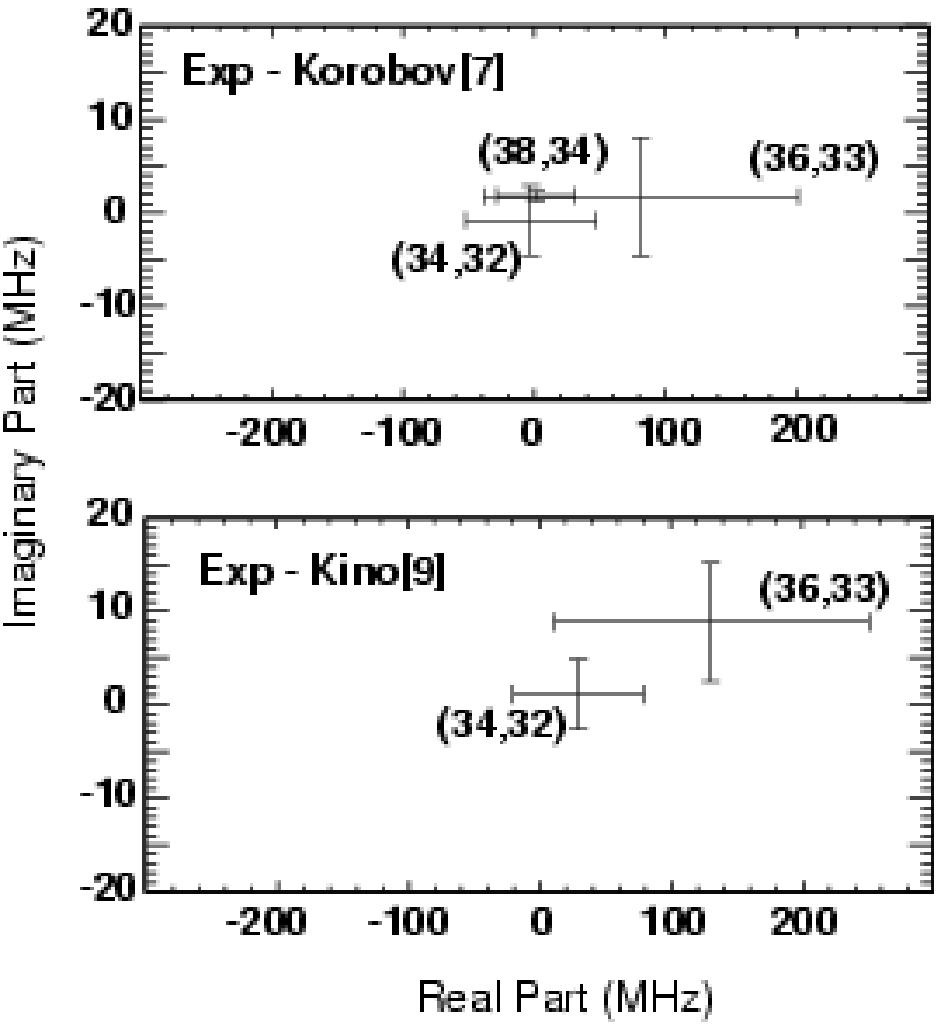}
\caption{\label{fig:complex}  
Precise comparison of the complex energies of 
three \pbhef\ states.
The difference of the experimental values 
from the two theories \cite{Korobov:03,Kino:03} are plotted in MHz.
The imaginary part plotted here is the 
half Auger width with negative sign, 
$-\frac{1}{2} (\gamma_{A}/2\pi)$.
The values used for the comparison of the real part are 
the transition frequencies from metastable states, where 
the theoretical uncertainty should be very small.
}
\end{center}
\end{figure}

Most of the experimental decay rates are roughly in agreement with
the calculations.
By using these results,
we were able to expand the recent precise experiment-theory
comparison of the level energies \cite{Hori:03} 
to the complex plane.
In Fig.~\ref{fig:complex}, the complex energy,
defined as Eq.~(\ref{complex_energy}),
was compared with values from 
two theories, \cite{Korobov:03} and \cite{Kino:03}.
For the imaginary part ({\it i.e.} the half Auger width)
of the three \pbhef\ states presented in Fig.~\ref{fig:complex},
no discrepancy 
greater than 10 MHz, was found. 
This limit is absolutely even more precise than
that obtained for the real part (i.e. the level energy).

A noteworthy state is  the (36, 32) of \pbhet.
Although an {\it ab-initio} calculation \cite{Korobov:97b} predicted 
a large anomaly, 
its decay rate was found to be non-anomalous.
Another calculation \cite{Kartavtsev:00} was done by introducing
continuum-coupled wave functions explicitly in the equation of motion,
and resulted in a non-anomalous value.
After the present experiment was finished
Korobov presented the results of a new calculation 
with the CCR method.
These are in good agreement with our experiment,
suggesting that the coupling with the continuum
is well represented by this new method.

As for the latest results obtained in this work,
two anomalous candidates have been found.
They were the (35, 32) state of \pbhef\ and the (34, 31) state of \pbhet, 
both of which have much faster decay rates 
than the typical values expected by Eq.~(\ref{auger_systematics}).
Theoretical Auger rates are also faster than the typical rates
for the (35, 32) state of \pbhef,
but there is a large discrepancy between the two CCR calculations
for the (34, 31) state of \pbhet. 
Korobov obtained an anomalous value 
in the old calculation \cite{Korobov:97b},
but his latest result \cite{Korobov:03} is non-anomalous,
while the value by Kino \cite{Kino:03} is anomalous.

\begin{figure}[!ht]
\begin{center}
\includegraphics[width=.4\linewidth]{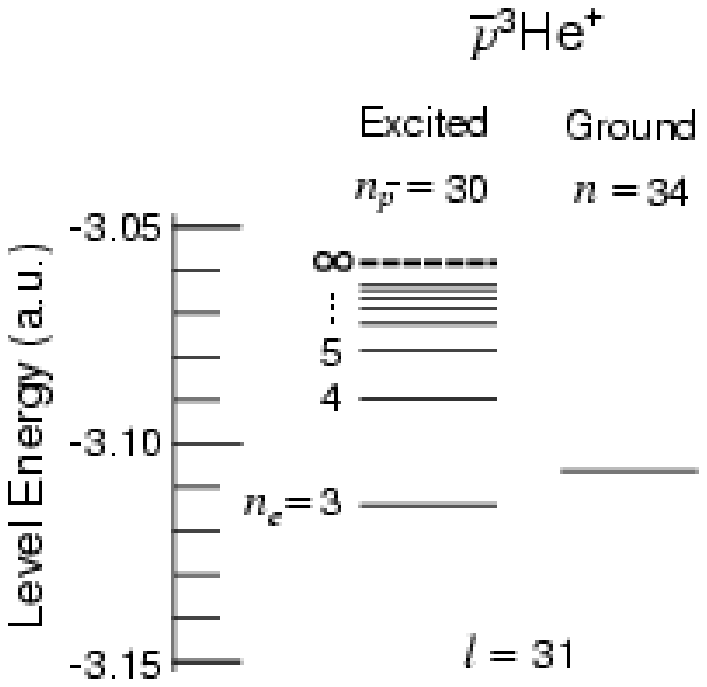}
\includegraphics[width=.4\linewidth]{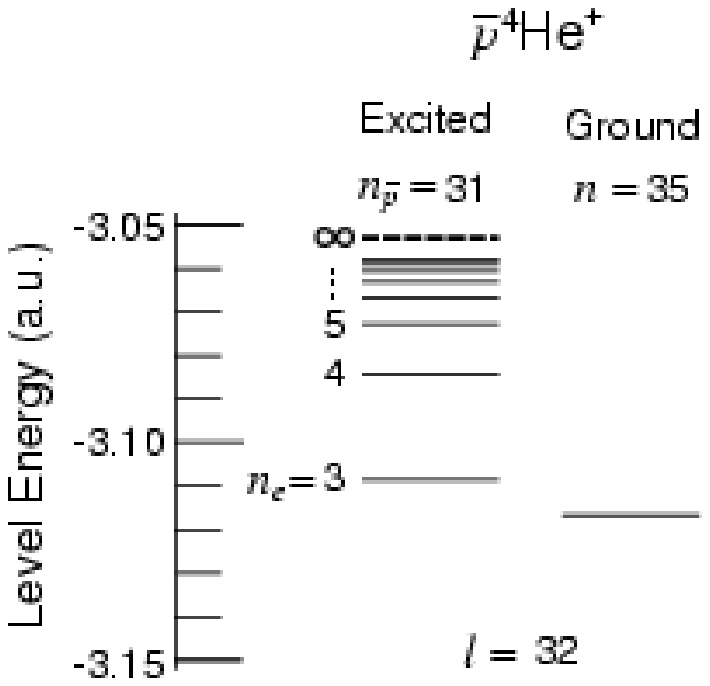}
\caption{\label{fig:partial}  Energy diagram of the ground-electron 
states (34, 31) of \pbhet\ and (35, 32) of \pbhef, 
and their neighbor electron-excited states.
The energy values of  the ground-electron states were obtained by
 \cite{Korobov:96}, and 
those of electron-excited states were by Eq.~(\ref{energy_ex}).
}
\end{center}
\end{figure}

These situations resemble the case of the (37, 33) state of \pbhef, 
the rate of which is supposed to be increased by a state
mixing with an electron-excited configuration having a nearby 
level energy.
Since an excited electron is far away from the nucleus 
compared to the 
distance of the other two particles, 
the energy of the electron-excited states can be approximated
by the following formula \cite{Kartavtsev:00,Yamaguchi:02},
\begin{equation}
E= -\frac{m_{\bar{p}}m_{\alpha}}{m_{\bar{p}}+m_{\alpha}} \frac{Z^2}{2n_{\bar{p}}
^2}-\frac{1}{2n_e^2},
\label{energy_ex}
\end{equation}
in atomic units. Here $Z = 2$ is the atomic number of helium.
As are illustrated in Fig.~\ref{fig:partial}, 
the configurations 
$(31, 30)_{\bar{p}} \otimes (3d)_e$ for the (35, 32) state of \pbhef, and 
$(30, 29)_{\bar{p}} \otimes (3d)_e$ for the (34, 31) state of \pbhet\ 
can be the reason for the anomalies in the same way.
For the latter case, the possibility of 
the influence by an electron-excited state 
was pointed out by Kartavtsev \cite{Kartavtsev:00}.

There is still a possibility that the rate enhancements were 
caused by collisions,
since we do not know the experimental decay rates 
of the atom isolated in vacuum.
In any case, studying these states both experimentally and theoretically 
should be a good test for understanding 
the continuum-coupled three-body system.

\section{Acknowledgement}

We are grateful to the CERN PS division for their help,
V.I. Korobov, D.D. Bakalov and Y. Kino for useful discussions.
This work was supported by the Grant-in-Aid for Specially Promoted
Research
(Grant No. 15002005) of MEXT, Japan,
and the Hungarian Scientific Research Fund (Grant Nos. OTKA T033079
and TeT-Jap-4/00).


\bibliography{ps205}

\newcommand{\SortNoop}[1]{} \newcommand{\OneLetter}[1]{#1}
  \newcommand{\SwapArgs}[2]{#2#1}
\begin{thebibliography}{23}
\expandafter\ifx\csname natexlab\endcsname\relax\def\natexlab#1{#1}\fi
\expandafter\ifx\csname bibnamefont\endcsname\relax
  \def\bibnamefont#1{#1}\fi
\expandafter\ifx\csname bibfnamefont\endcsname\relax
  \def\bibfnamefont#1{#1}\fi
\expandafter\ifx\csname citenamefont\endcsname\relax
  \def\citenamefont#1{#1}\fi
\expandafter\ifx\csname url\endcsname\relax
  \def\url#1{\texttt{#1}}\fi
\expandafter\ifx\csname urlprefix\endcsname\relax\def\urlprefix{URL }\fi
\providecommand{\bibinfo}[2]{#2}
\providecommand{\eprint}[2][]{\url{#2}}

\bibitem[{\citenamefont{Iwasaki et~al.}(1991)\citenamefont{Iwasaki, Nakamura,
  Shigaki, Shimizu, Tamura, Ishikawa, Hayano, Takada, Widmann, Outa, Aoki,
  Kitching, and Yamazaki}}]{Iwasaki:91}
\bibinfo{author}{\bibfnamefont{M.}~\bibnamefont{Iwasaki}},
  \bibinfo{author}{\bibfnamefont{S.~N.} \bibnamefont{Nakamura}},
  \bibinfo{author}{\bibfnamefont{K.}~\bibnamefont{Shigaki}},
  \bibinfo{author}{\bibfnamefont{Y.}~\bibnamefont{Shimizu}},
  \bibinfo{author}{\bibfnamefont{H.}~\bibnamefont{Tamura}},
  \bibinfo{author}{\bibfnamefont{T.}~\bibnamefont{Ishikawa}},
  \bibinfo{author}{\bibfnamefont{R.~S.} \bibnamefont{Hayano}},
  \bibinfo{author}{\bibfnamefont{E.}~\bibnamefont{Takada}},
  \bibinfo{author}{\bibfnamefont{E.}~\bibnamefont{Widmann}},
  \bibinfo{author}{\bibfnamefont{H.}~\bibnamefont{Outa}},
  \bibinfo{author}{\bibfnamefont{M.}~\bibnamefont{Aoki}},
  \bibinfo{author}{\bibfnamefont{P.}~\bibnamefont{Kitching}}, \bibnamefont{and}
  \bibinfo{author}{\bibfnamefont{T.}~\bibnamefont{Yamazaki}},
  \bibinfo{journal}{Phys. Rev. Lett.} \textbf{\bibinfo{volume}{67}},
  \bibinfo{pages}{1246} (\bibinfo{year}{1991}).

\bibitem[{\citenamefont{Yamazaki et~al.}(1993)\citenamefont{Yamazaki, Widmann,
  Hayano, Iwasaki, Nakamura, Shigaki, Hartmann, Daniel, von Egidy, Hofmann,
  Kim, and Eades}}]{Yamazaki:93}
\bibinfo{author}{\bibfnamefont{T.}~\bibnamefont{Yamazaki}},
  \bibinfo{author}{\bibfnamefont{E.}~\bibnamefont{Widmann}},
  \bibinfo{author}{\bibfnamefont{R.~S.} \bibnamefont{Hayano}},
  \bibinfo{author}{\bibfnamefont{M.}~\bibnamefont{Iwasaki}},
  \bibinfo{author}{\bibfnamefont{S.~N.} \bibnamefont{Nakamura}},
  \bibinfo{author}{\bibfnamefont{K.}~\bibnamefont{Shigaki}},
  \bibinfo{author}{\bibfnamefont{F.~J.} \bibnamefont{Hartmann}},
  \bibinfo{author}{\bibfnamefont{H.}~\bibnamefont{Daniel}},
  \bibinfo{author}{\bibfnamefont{T.}~\bibnamefont{von Egidy}},
  \bibinfo{author}{\bibfnamefont{P.}~\bibnamefont{Hofmann}},
  \bibinfo{author}{\bibfnamefont{Y.-S.} \bibnamefont{Kim}}, \bibnamefont{and}
  \bibinfo{author}{\bibfnamefont{J.}~\bibnamefont{Eades}},
  \bibinfo{journal}{Nature} \textbf{\bibinfo{volume}{361}},
  \bibinfo{pages}{238} (\bibinfo{year}{1993}).

\bibitem[{\citenamefont{Yamazaki et~al.}(2002)\citenamefont{Yamazaki, Morita,
  Hayano, Widmann, and Eades}}]{Yamazaki:02}
\bibinfo{author}{\bibfnamefont{T.}~\bibnamefont{Yamazaki}},
  \bibinfo{author}{\bibfnamefont{N.}~\bibnamefont{Morita}},
  \bibinfo{author}{\bibfnamefont{R.~S.} \bibnamefont{Hayano}},
  \bibinfo{author}{\bibfnamefont{E.}~\bibnamefont{Widmann}}, \bibnamefont{and}
  \bibinfo{author}{\bibfnamefont{J.}~\bibnamefont{Eades}},
  \bibinfo{journal}{Physics Reports} \textbf{\bibinfo{volume}{366}},
  \bibinfo{pages}{183} (\bibinfo{year}{2002}).

\bibitem[{\citenamefont{Hori et~al.}(2003)\citenamefont{Hori, Eades, Hayano,
  Ishikawa, Pirkl, Widmann, Yamaguchi, Torii, Juh\'{a}sz, Horv\'{a}th, and
  Yamazaki}}]{Hori:03}
\bibinfo{author}{\bibfnamefont{M.}~\bibnamefont{Hori}},
  \bibinfo{author}{\bibfnamefont{J.}~\bibnamefont{Eades}},
  \bibinfo{author}{\bibfnamefont{R.~S.} \bibnamefont{Hayano}},
  \bibinfo{author}{\bibfnamefont{T.}~\bibnamefont{Ishikawa}},
  \bibinfo{author}{\bibfnamefont{W.}~\bibnamefont{Pirkl}},
  \bibinfo{author}{\bibfnamefont{E.}~\bibnamefont{Widmann}},
  \bibinfo{author}{\bibfnamefont{H.}~\bibnamefont{Yamaguchi}},
  \bibinfo{author}{\bibfnamefont{H.~A.} \bibnamefont{Torii}},
  \bibinfo{author}{\bibfnamefont{B.}~\bibnamefont{Juh\'{a}sz}},
  \bibinfo{author}{\bibfnamefont{D.}~\bibnamefont{Horv\'{a}th}},
  \bibnamefont{and} \bibinfo{author}{\bibfnamefont{T.}~\bibnamefont{Yamazaki}},
  \bibinfo{journal}{Phys. Rev. Lett.} \textbf{\bibinfo{volume}{91}},
  \bibinfo{pages}{123401} (\bibinfo{year}{2003}).

\bibitem[{\citenamefont{Ko{\OneLetter{ro}}bov}(1996)}]{Korobov:96}
\bibinfo{author}{\bibfnamefont{V.~I.} \bibnamefont{Ko{\OneLetter{ro}}bov}},
  \bibinfo{journal}{Phys. Rev. A} \textbf{\bibinfo{volume}{54}},
  \bibinfo{pages}{R1749} (\bibinfo{year}{1996}).

\bibitem[{\citenamefont{Ko{\OneLetter{ro}}bov and Bakalov}(1997)}]{Korobov:97a}
\bibinfo{author}{\bibfnamefont{V.~I.} \bibnamefont{Ko{\OneLetter{ro}}bov}}
  \bibnamefont{and} \bibinfo{author}{\bibfnamefont{D.~D.}
  \bibnamefont{Bakalov}}, \bibinfo{journal}{Phys. Rev. Lett.}
  \textbf{\bibinfo{volume}{79}}, \bibinfo{pages}{3379} (\bibinfo{year}{1997}).

\bibitem[{\citenamefont{Ko{\OneLetter{ro}}bov}(2003)}]{Korobov:03}
\bibinfo{author}{\bibfnamefont{V.~I.} \bibnamefont{Ko{\OneLetter{ro}}bov}},
  \bibinfo{journal}{Phys. Rev. A} \textbf{\bibinfo{volume}{67}},
  \bibinfo{pages}{062501} (\bibinfo{year}{2003}), \bibinfo{note}{Erratum 
\textbf{68} 019902 (2003)}.

\bibitem[{\citenamefont{Kino et~al.}(2001)\citenamefont{Kino, Yamanaka,
  Kamimura, Froelich, and Kudo}}]{Kino:01}
\bibinfo{author}{\bibfnamefont{Y.}~\bibnamefont{Kino}},
  \bibinfo{author}{\bibfnamefont{N.}~\bibnamefont{Yamanaka}},
  \bibinfo{author}{\bibfnamefont{M.}~\bibnamefont{Kamimura}},
  \bibinfo{author}{\bibfnamefont{P.}~\bibnamefont{Froelich}}, \bibnamefont{and}
  \bibinfo{author}{\bibfnamefont{H.}~\bibnamefont{Kudo}},
  \bibinfo{journal}{Hyperfine Intractions} \textbf{\bibinfo{volume}{138}},
  \bibinfo{pages}{179} (\bibinfo{year}{2001}).

\bibitem[{\citenamefont{Kino et~al.}()\citenamefont{Kino, Kamimura, and
  Kudo}}]{Kino:03}
\bibinfo{author}{\bibfnamefont{Y.}~\bibnamefont{Kino}},
  \bibinfo{author}{\bibfnamefont{M.}~\bibnamefont{Kamimura}}, \bibnamefont{and}
  \bibinfo{author}{\bibfnamefont{H.}~\bibnamefont{Kudo}},
  \bibinfo{journal}{Nucl. Instrum. Methods Phys. Res., Sect. B},
  \bibinfo{note}{in press}.

\bibitem[{\citenamefont{Russell}(1970)}]{Russell:70c}
\bibinfo{author}{\bibfnamefont{J.~E.} \bibnamefont{Russell}},
  \bibinfo{journal}{Phys. Rev. A} \textbf{\bibinfo{volume}{1}},
  \bibinfo{pages}{742} (\bibinfo{year}{1970}).

\bibitem[{\citenamefont{Kartavtsev et~al.}(2000)\citenamefont{Kartavtsev,
  Monakhov, and Fedotov}}]{Kartavtsev:00}
\bibinfo{author}{\bibfnamefont{O.}~\bibnamefont{Kartavtsev}},
  \bibinfo{author}{\bibfnamefont{D.}~\bibnamefont{Monakhov}}, \bibnamefont{and}
  \bibinfo{author}{\bibfnamefont{S.}~\bibnamefont{Fedotov}},
  \bibinfo{journal}{Phys. Rev. A} \textbf{\bibinfo{volume}{61}},
  \bibinfo{pages}{062507} (\bibinfo{year}{2000}), \bibinfo{note}{Erratum 
   \textbf{63},  019901(E) (2000)}.


\bibitem[{\citenamefont{Yamazaki and Ohtsuki}(1992)}]{Yamazaki:92}
\bibinfo{author}{\bibfnamefont{T.}~\bibnamefont{Yamazaki}} \bibnamefont{and}
  \bibinfo{author}{\bibfnamefont{K.}~\bibnamefont{Ohtsuki}},
  \bibinfo{journal}{Phys. Rev. A} \textbf{\bibinfo{volume}{45}},
  \bibinfo{pages}{7782} (\bibinfo{year}{1992}); \bibinfo{note}{ K. Ohtsuki,
  unpublished theoretical data (1992)}.

\bibitem[{\citenamefont{Ko{\OneLetter{ro}}bov and
  Shimamura}(1997)}]{Korobov:97b}
\bibinfo{author}{\bibfnamefont{V.~I.} \bibnamefont{Ko{\OneLetter{ro}}bov}}
  \bibnamefont{and}
  \bibinfo{author}{\bibfnamefont{I.}~\bibnamefont{Shimamura}},
  \bibinfo{journal}{Phys. Rev. A} \textbf{\bibinfo{volume}{56}},
  \bibinfo{pages}{4587} (\bibinfo{year}{1997}).

\bibitem[{\citenamefont{Hori et~al.}(1998)\citenamefont{Hori, Torii, Hayano,
  Ishikawa, Maas, Tamura, Ketzer, Hartmann, Pohl, Maierl, Hasinoff, von Egidy,
  Kumakura, Morita, Sugai, Horv\'ath, Widmann, Eades, and Yamazaki}}]{Hori:98}
\bibinfo{author}{\bibfnamefont{M.}~\bibnamefont{Hori}},
  \bibinfo{author}{\bibfnamefont{H.~A.} \bibnamefont{Torii}},
  \bibinfo{author}{\bibfnamefont{R.~S.} \bibnamefont{Hayano}},
  \bibinfo{author}{\bibfnamefont{T.}~\bibnamefont{Ishikawa}},
  \bibinfo{author}{\bibfnamefont{F.~E.} \bibnamefont{Maas}},
  \bibinfo{author}{\bibfnamefont{H.}~\bibnamefont{Tamura}},
  \bibinfo{author}{\bibfnamefont{B.}~\bibnamefont{Ketzer}},
  \bibinfo{author}{\bibfnamefont{F.~J.} \bibnamefont{Hartmann}},
  \bibinfo{author}{\bibfnamefont{R.}~\bibnamefont{Pohl}},
  \bibinfo{author}{\bibfnamefont{C.}~\bibnamefont{Maierl}},
  \bibinfo{author}{\bibfnamefont{M.}~\bibnamefont{Hasinoff}},
  \bibinfo{author}{\bibfnamefont{T.}~\bibnamefont{von Egidy}},
  \bibinfo{author}{\bibfnamefont{M.}~\bibnamefont{Kumakura}},
  \bibinfo{author}{\bibfnamefont{N.}~\bibnamefont{Morita}},
  \bibinfo{author}{\bibfnamefont{I.}~\bibnamefont{Sugai}},
  \bibinfo{author}{\bibfnamefont{D.}~\bibnamefont{Horv\'ath}},
  \bibinfo{author}{\bibfnamefont{E.}~\bibnamefont{Widmann}},
  \bibinfo{author}{\bibfnamefont{J.}~\bibnamefont{Eades}}, \bibnamefont{and}
  \bibinfo{author}{\bibfnamefont{T.}~\bibnamefont{Yamazaki}},
  \bibinfo{journal}{Phys. Rev. A} \textbf{\bibinfo{volume}{57}},
  \bibinfo{pages}{1698} (\bibinfo{year}{1998}), \bibinfo{note}{Erratum \textbf{58} 1612 (1998)}.

\bibitem[{\citenamefont{Yamaguchi et~al.}(2002)\citenamefont{Yamaguchi,
  Ishikawa, Sakaguchi, Widmann, Eades, Hayano, Hori, Torii, Juh\'{a}sz,
  Horv\'{a}th, and Yamazaki}}]{Yamaguchi:02}
\bibinfo{author}{\bibfnamefont{H.}~\bibnamefont{Yamaguchi}},
  \bibinfo{author}{\bibfnamefont{T.}~\bibnamefont{Ishikawa}},
  \bibinfo{author}{\bibfnamefont{J.}~\bibnamefont{Sakaguchi}},
  \bibinfo{author}{\bibfnamefont{E.}~\bibnamefont{Widmann}},
  \bibinfo{author}{\bibfnamefont{J.}~\bibnamefont{Eades}},
  \bibinfo{author}{\bibfnamefont{R.~S.} \bibnamefont{Hayano}},
  \bibinfo{author}{\bibfnamefont{M.}~\bibnamefont{Hori}},
  \bibinfo{author}{\bibfnamefont{H.~A.} \bibnamefont{Torii}},
  \bibinfo{author}{\bibfnamefont{B.}~\bibnamefont{Juh\'{a}sz}},
  \bibinfo{author}{\bibfnamefont{D.}~\bibnamefont{Horv\'{a}th}},
  \bibnamefont{and} \bibinfo{author}{\bibfnamefont{T.}~\bibnamefont{Yamazaki}},
  \bibinfo{journal}{Phys. Rev. A} \textbf{\bibinfo{volume}{66}},
  \bibinfo{pages}{022504} (\bibinfo{year}{2002}).

\bibitem[{\citenamefont{Lombardi et~al.}(2001)\citenamefont{Lombardi, Pirkl,
  and Bylinsky}}]{Lombardi:01}
\bibinfo{author}{\bibfnamefont{A.~M.} \bibnamefont{Lombardi}},
  \bibinfo{author}{\bibfnamefont{W.}~\bibnamefont{Pirkl}}, \bibnamefont{and}
  \bibinfo{author}{\bibfnamefont{Y.}~\bibnamefont{Bylinsky}}, in
  \emph{\bibinfo{booktitle}{Proceedings of the 2001 Particle Accelerator
  Conference, Chicago}} (\bibinfo{publisher}{IEEE, Piscataway, NJ},
  \bibinfo{year}{2001}), pp. \bibinfo{pages}{585--587}.

\bibitem[{\citenamefont{Hori et~al.}(2001)\citenamefont{Hori, Eades, Widmann,
  Yamaguchi, Sakaguchi, Ishikawa, Hayano, Torii, Juh\'asz, Horv\'ath, and
  Yamazaki}}]{Hori:01}
\bibinfo{author}{\bibfnamefont{M.}~\bibnamefont{Hori}},
  \bibinfo{author}{\bibfnamefont{J.}~\bibnamefont{Eades}},
  \bibinfo{author}{\bibfnamefont{E.}~\bibnamefont{Widmann}},
  \bibinfo{author}{\bibfnamefont{H.}~\bibnamefont{Yamaguchi}},
  \bibinfo{author}{\bibfnamefont{J.}~\bibnamefont{Sakaguchi}},
  \bibinfo{author}{\bibfnamefont{T.}~\bibnamefont{Ishikawa}},
  \bibinfo{author}{\bibfnamefont{R.~S.} \bibnamefont{Hayano}},
  \bibinfo{author}{\bibfnamefont{H.~A.} \bibnamefont{Torii}},
  \bibinfo{author}{\bibfnamefont{B.}~\bibnamefont{Juh\'asz}},
  \bibinfo{author}{\bibfnamefont{D.}~\bibnamefont{Horv\'ath}},
  \bibnamefont{and} \bibinfo{author}{\bibfnamefont{T.}~\bibnamefont{Yamazaki}},
  \bibinfo{journal}{Phys. Rev. Lett.} \textbf{\bibinfo{volume}{87}},
  \bibinfo{pages}{093401} (\bibinfo{year}{2001}).

\bibitem[{\citenamefont{Morita et~al.}(1993)\citenamefont{Morita, Ohtsuki, and
  Yamazaki}}]{Morita:93}
\bibinfo{author}{\bibfnamefont{N.}~\bibnamefont{Morita}},
  \bibinfo{author}{\bibfnamefont{K.}~\bibnamefont{Ohtsuki}}, \bibnamefont{and}
  \bibinfo{author}{\bibfnamefont{T.}~\bibnamefont{Yamazaki}},
  \bibinfo{journal}{Nucl. Instrum. Methods Phys. Res., Sect. A}
  \textbf{\bibinfo{volume}{330}}, \bibinfo{pages}{439} (\bibinfo{year}{1993}).

\bibitem[{\citenamefont{Morita et~al.}(1994)\citenamefont{Morita, Kumakura,
  Yamazaki, Widmann, Masuda, Sugai, Hayano, Maas, Torii, Hartmann, Daniel, von
  Egidy, Ketzer, M{\"u}ller, Schmid, Horv\'ath, and Eades}}]{Morita:94}
\bibinfo{author}{\bibfnamefont{N.}~\bibnamefont{Morita}},
  \bibinfo{author}{\bibfnamefont{M.}~\bibnamefont{Kumakura}},
  \bibinfo{author}{\bibfnamefont{T.}~\bibnamefont{Yamazaki}},
  \bibinfo{author}{\bibfnamefont{E.}~\bibnamefont{Widmann}},
  \bibinfo{author}{\bibfnamefont{H.}~\bibnamefont{Masuda}},
  \bibinfo{author}{\bibfnamefont{I.}~\bibnamefont{Sugai}},
  \bibinfo{author}{\bibfnamefont{R.~S.} \bibnamefont{Hayano}},
  \bibinfo{author}{\bibfnamefont{F.~E.} \bibnamefont{Maas}},
  \bibinfo{author}{\bibfnamefont{H.~A.} \bibnamefont{Torii}},
  \bibinfo{author}{\bibfnamefont{F.~J.} \bibnamefont{Hartmann}},
  \bibinfo{author}{\bibfnamefont{H.}~\bibnamefont{Daniel}},
  \bibinfo{author}{\bibfnamefont{T.}~\bibnamefont{von Egidy}},
  \bibinfo{author}{\bibfnamefont{B.}~\bibnamefont{Ketzer}},
  \bibinfo{author}{\bibfnamefont{W.}~\bibnamefont{M{\"u}ller}},
  \bibinfo{author}{\bibfnamefont{W.}~\bibnamefont{Schmid}},
  \bibinfo{author}{\bibfnamefont{D.}~\bibnamefont{Horv\'ath}},
  \bibnamefont{and} \bibinfo{author}{\bibfnamefont{J.}~\bibnamefont{Eades}},
  \bibinfo{journal}{Phys. Rev. Lett.} \textbf{\bibinfo{volume}{72}},
  \bibinfo{pages}{1180} (\bibinfo{year}{1994}).

\bibitem[{\citenamefont{Niestroj et~al.}(1996)\citenamefont{Niestroj, Hartmann,
  Daniel, Ketzer, von Egidy, Maas, Hayano, Ishikawa, Tamura, Torii, Morita,
  Yamazaki, Sugai, Nakayoshi, Horv\'ath, Eades, and Widmann}}]{Niestroj:96}
\bibinfo{author}{\bibfnamefont{A.}~\bibnamefont{Niestroj}},
  \bibinfo{author}{\bibfnamefont{F.~J.} \bibnamefont{Hartmann}},
  \bibinfo{author}{\bibfnamefont{H.}~\bibnamefont{Daniel}},
  \bibinfo{author}{\bibfnamefont{B.}~\bibnamefont{Ketzer}},
  \bibinfo{author}{\bibfnamefont{T.}~\bibnamefont{von Egidy}},
  \bibinfo{author}{\bibfnamefont{F.~E.} \bibnamefont{Maas}},
  \bibinfo{author}{\bibfnamefont{R.~S.} \bibnamefont{Hayano}},
  \bibinfo{author}{\bibfnamefont{T.}~\bibnamefont{Ishikawa}},
  \bibinfo{author}{\bibfnamefont{H.}~\bibnamefont{Tamura}},
  \bibinfo{author}{\bibfnamefont{H.~A.} \bibnamefont{Torii}},
  \bibinfo{author}{\bibfnamefont{N.}~\bibnamefont{Morita}},
  \bibinfo{author}{\bibfnamefont{T.}~\bibnamefont{Yamazaki}},
  \bibinfo{author}{\bibfnamefont{I.}~\bibnamefont{Sugai}},
  \bibinfo{author}{\bibfnamefont{K.}~\bibnamefont{Nakayoshi}},
  \bibinfo{author}{\bibfnamefont{D.}~\bibnamefont{Horv\'ath}},
  \bibinfo{author}{\bibfnamefont{J.}~\bibnamefont{Eades}}, \bibnamefont{and}
  \bibinfo{author}{\bibfnamefont{E.}~\bibnamefont{Widmann}},
  \bibinfo{journal}{Nucl. Instrum. Methods Phys. Res., Sect. A}
  \textbf{\bibinfo{volume}{373}}, \bibinfo{pages}{411} (\bibinfo{year}{1996}).

\bibitem[{\citenamefont{Widmann et~al.}(1997)\citenamefont{Widmann, Eades,
  Yamazaki, Torii, Hayano, Hori, Ishikawa, Kumakura, Morita, Sugai, Hartmann,
  von Egidy, Ketzer, Maierl, Pohl, and Horv\'ath}}]{Widmann:97}
\bibinfo{author}{\bibfnamefont{E.}~\bibnamefont{Widmann}},
  \bibinfo{author}{\bibfnamefont{J.}~\bibnamefont{Eades}},
  \bibinfo{author}{\bibfnamefont{T.}~\bibnamefont{Yamazaki}},
  \bibinfo{author}{\bibfnamefont{H.~A.} \bibnamefont{Torii}},
  \bibinfo{author}{\bibfnamefont{R.~S.} \bibnamefont{Hayano}},
  \bibinfo{author}{\bibfnamefont{M.}~\bibnamefont{Hori}},
  \bibinfo{author}{\bibfnamefont{T.}~\bibnamefont{Ishikawa}},
  \bibinfo{author}{\bibfnamefont{M.}~\bibnamefont{Kumakura}},
  \bibinfo{author}{\bibfnamefont{N.}~\bibnamefont{Morita}},
  \bibinfo{author}{\bibfnamefont{I.}~\bibnamefont{Sugai}},
  \bibinfo{author}{\bibfnamefont{F.}~\bibnamefont{Hartmann}},
  \bibinfo{author}{\bibfnamefont{T.}~\bibnamefont{von Egidy}},
  \bibinfo{author}{\bibfnamefont{B.}~\bibnamefont{Ketzer}},
  \bibinfo{author}{\bibfnamefont{C.}~\bibnamefont{Maierl}},
  \bibinfo{author}{\bibfnamefont{R.}~\bibnamefont{Pohl}}, \bibnamefont{and}
  \bibinfo{author}{\bibfnamefont{D.}~\bibnamefont{Horv\'ath}},
  \bibinfo{journal}{Phys. Lett. B} \textbf{\bibinfo{volume}{404}},
  \bibinfo{pages}{15} (\bibinfo{year}{1997}).

\bibitem[{\citenamefont{Widmann et~al.}(2002)\citenamefont{Widmann, Eades,
  Ishikawa, Sakaguchi, Tasaki, Yamaguchi, Hayano, Hori, Torii, Juh\'{a}sz,
  Horv\'{a}th, and Yamazaki}}]{Widmann:02}
\bibinfo{author}{\bibfnamefont{E.}~\bibnamefont{Widmann}},
  \bibinfo{author}{\bibfnamefont{J.}~\bibnamefont{Eades}},
  \bibinfo{author}{\bibfnamefont{T.}~\bibnamefont{Ishikawa}},
  \bibinfo{author}{\bibfnamefont{J.}~\bibnamefont{Sakaguchi}},
  \bibinfo{author}{\bibfnamefont{T.}~\bibnamefont{Tasaki}},
  \bibinfo{author}{\bibfnamefont{H.}~\bibnamefont{Yamaguchi}},
  \bibinfo{author}{\bibfnamefont{R.~S.} \bibnamefont{Hayano}},
  \bibinfo{author}{\bibfnamefont{M.}~\bibnamefont{Hori}},
  \bibinfo{author}{\bibfnamefont{H.~A.} \bibnamefont{Torii}},
  \bibinfo{author}{\bibfnamefont{B.}~\bibnamefont{Juh\'{a}sz}},
  \bibinfo{author}{\bibfnamefont{D.}~\bibnamefont{Horv\'{a}th}},
  \bibnamefont{and} \bibinfo{author}{\bibfnamefont{T.}~\bibnamefont{Yamazaki}},
  \bibinfo{journal}{Phys. Rev. Lett.} \textbf{\bibinfo{volume}{89}},
  \bibinfo{pages}{243402} (\bibinfo{year}{2002}).

\bibitem[{\citenamefont{Bakalov and Korobov}(1998)}]{Bakalov:98}
\bibinfo{author}{\bibfnamefont{D.}~\bibnamefont{Bakalov}} \bibnamefont{and}
  \bibinfo{author}{\bibfnamefont{V.~I.} \bibnamefont{Korobov}},
  \bibinfo{journal}{Phys. Rev. A} \textbf{\bibinfo{volume}{57}},
  \bibinfo{pages}{1662} (\bibinfo{year}{1998}).

\end{thebibliography}

\end{document}